\documentclass[11pt,showpacs,preprintnumbers,amsmath,amssymb]{revtex4}

\usepackage[dvips]{graphicx}
\usepackage{epsfig}

\begin{document}

\title{
Staggered dimer order in 
S=$\frac{1}{2}$ quantum spin ladder system
with four spin exchange
}

\author{ Keigo Hijii$^1$, Shaojin Qin$^2$, and Kiyohide Nomura$^1$ }
\affiliation{
 $^1$ {\it Department of Physics,Kyushu University, Fukuoka 812-8581, Japan}
\\
 $^2$ {\it Institute of Theoretical Physics, P O Box 2735, Beijing, 100871, P R China}
}


\large

\begin{abstract}

We study the S=$\frac{1}{2}$ quantum spin ladder system
with the four-spin exchange,
using density matrix renormalization group method
and an exact diagonalization method.
Recently, the phase transition in this system
and its universality class are studied.
But there remain controversies
whether the phase transition is second order type or the other type
and the nature of order parameter.
There are arguments that the massless phase appears.
But this does not agree with our previous result.
Analyzing DMRG data,
we try a new approach in order to determine a phase which appears after
the phase transition.
We find that the edge state appears in the open boundary condition,
investigating excitation energies 
of states with higher magnetizations.

\end{abstract}

\maketitle

\section{INTRODUCTION}

S=1/2 quantum spin two-leg ladder systems have been studied
from both theoretical and experimental points of view \cite{D-R}.
These systems are related to high-T$_{c}$ superconductors
and Haldane's conjecture\cite{Haldane}.
For example, Dagotto {\it et al.} suggested that 
the superconductivity occurs in hole-doped ladder systems 
\cite{D-R-S}.

Usually, two spin exchanges have been discussed 
in quantum spin systems.
By the way, spin exchange interactions originate from electron exchanges.
For example, Heisenberg model is 
derived from the second order perturbation
in the strong coupling limit of the Hubbard model.
In contrast, higher order perturbations 
give many body spin exchange interactions,
for example,
$\left( {\bf S}_i \cdot {\bf S}_j \right) 
\left( {\bf S}_k \cdot {\bf S}_l \right)$.
These type interactions in two leg ladder systems are discussed
by Nersesyan {\it et al}\cite{N-T}.
Investigating correlation functions, they referred to
level $k=2$ $SU(2)$ Wess-Zumino-Witten (WZW) model or $SU(2)$ $c=3/2$ 
conformal field theory (CFT)).

\begin{figure}
\begin{center}
 \epsfig{file=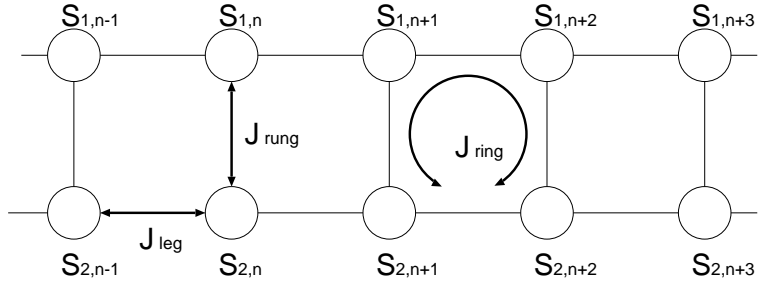,width=10cm}
 \caption{
Two-leg spin ladder with four spin exchange.
 }
\label{fig:4spin-ladder}
\end{center}
\end{figure}

Recently four spin cyclic exchanges attract attention
\cite{N-B-K-W-G,S-K-U,B-L-S-U,S-M-U,M-K-E-B-M-1,M-K-E-B-M-2,
C-H-A-P-F-M-C-F,H-K-W,S-H,N-T-H-O-S,H-N,M-V-M,L-S-T,H-H}. 
And from experiments,
it has been suggested that a four spin cyclic exchange interaction
plays an important role in some systems,
one dimensional two-leg ladder in La$_{6}$Ca$_{8}$Cu$_{24}$O$_{41}$
\cite{M-K-E-B-M-1,M-K-E-B-M-2}
and two dimensional square lattice in La$_{2}$CuO$_{4}$
\cite{C-H-A-P-F-M-C-F}.
The latter is known as a parent insulator of high-T$_{c}$
superconducting systems.
The importance of a cyclic four spin exchange in such systems
is pointed out by Honda {\it et al} \cite{H-K-W}.
In two-leg ladder systems (see Fig. \ref{fig:4spin-ladder}),
four spin exchanges play important roles. 
Sakai {\it et al} \cite{S-H} and Nakasu {\it et al} \cite{N-T-H-O-S}
studied this system with magnetizations in a magnetic field, 
in respect of the magnetization plateau and plateau-gapless transition.
This phase transition is the BKT type\cite{B-K-T}.

\begin{figure}
\begin{center}
 \epsfig{file=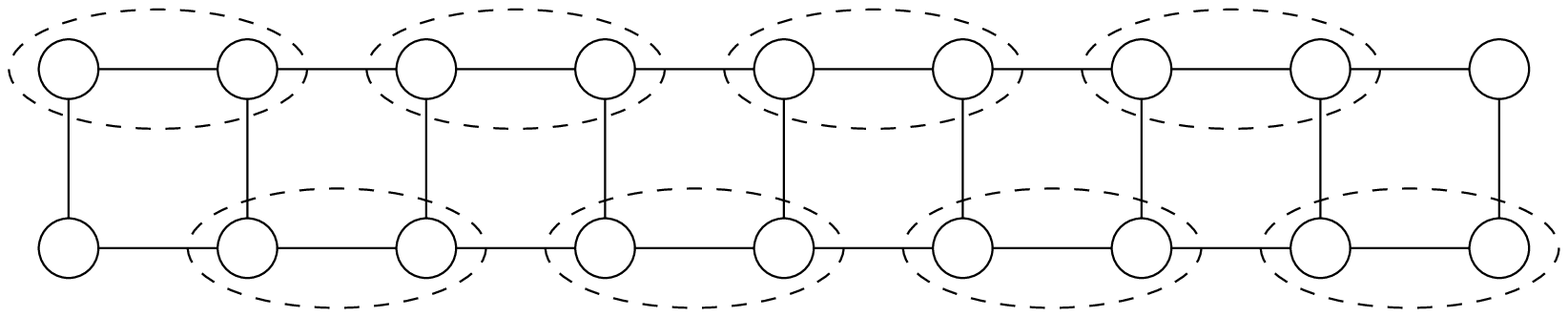,width=10cm}
 \caption{
A staggered dimer order state which consists of two spins
${\bf S}_{\alpha,i},{\bf S}_{\alpha,i+1}$
enclosed by a dotted line, in the open boundary condition.
This phase has an order parameter 
$\left\langle {\bf S}_{\alpha,i} \cdot {\bf S}_{\alpha,i+1} \right\rangle$
-$\left\langle {\bf S}_{\alpha,i} \cdot {\bf S}_{\alpha,i+1} \right\rangle$
 }
\label{fig:vbs}
\end{center}
\end{figure}

In our previous paper,
we numerically studied s=1/2 spin ladder system under zero field,
and we found a second order phase transition \cite{H-N}.
We studied the critical behavior of this system based on 
conformal field theory \cite{gins}.
We calculated the central charge and 
one of the scaling dimension numerically,
and we obtained $c \simeq \frac{3}{2}$, $x \simeq \frac{3}{8}$. 
Therefore this phase transition is described as 
the $c=\frac{3}{2}$ CFT or 
the $k=2$ $SU(2)$ WZW model.
This is a phase transition with ${\bf Z}_2$ symmetry breaking.
We discussed that the ordered state has the translational symmetry breaking
and the rung-parity symmetry breaking, which supports the
staggered dimer order (see Fig. \ref{fig:vbs}). 
Recently this result is obtained by M\"uller {\it et al} \cite{M-V-M}
in analytically.
They used bosonization technique and its Majorana representation
in the weak coupling limit.
With the help of Pad\'e-approximation in the strong coupling limit,
they obtained the critical exponent for the energy gap,
$\eta \simeq 1$ numerically.
This corresponds with our result\cite{H-N}.
L\"auchli {\it et al} \cite{L-S-T} discussed
a phase diagram for this model.
They calculated staggered dimer structure factor 
with exact diagonalization and a local 
$\left< {\bf S}_{\alpha,i} \cdot {\bf S}_{\alpha,i+1} \right>$ 
expectation value on one of the two leg in the dimer long range order
phase with 
the density matrix renormalization group (DMRG) method.
As a result, they conclude that 
there is a phase transition between 
a rung-singlet phase and a staggered dimer phase.  
It is clear and understandable to see the order parameter directly
as L\"auchli.
But it is needed to inspect the accuracy of numerical calculation and 
finite size effects since the correlation length is large. 
In fact, as we can see later, we need a careful analysis 
on the DMRG data.

However these results do not correspond to Honda {\it et al} \cite{H-H}.
They insisted that a massless phase appears 
in the strong four-spin coupling region.
Honda {\it et al} investigated a spin gap and a spin-pair correlation function
using DMRG method \cite{DMRG}.
They found that at $J_{rung}=J_{leg}=1$ and $J_{ring} \simeq 0.3$
 a spin gap between the singlet ($S^z_{tot}=0$) 
and the triplet ($S^z_{tot}=0,\pm 1$) vanishes.

Hikihara {\it et al} \cite{H-M-H} discussed the self-duality 
and the chirality using Runge-Lenz vector.  
And they referred to an incommensurate character.
They derived the duality transformation which leaves the form
of the Hamiltonian unchanged.
This self-dual line is same to an exact line 
which was found by Kolezhuk et al \cite{K-M}.
With the matrix product method,
one can exactly solve this model on this line
\cite{K-M}.
The exact ground state is a product of rung-singlets
on this line. 

Hikihara {\it et al} calculated spin gaps, a triplet excitation and
a quintet excitation,
using DMRG in open boundary condition.
They found out that both gaps are quite small in 
$J_{ring}$ large region.
They found that triplet gaps are exactly zero 
within numerical accuracy in some system sizes,
but quintet gaps seem to be finite.
They said that it is very difficult to estimate the critical point 
accurately due to the very slow vanishing of the spin gaps around
the phase transition point.

The extended massless phase is not consistent with some gapful phase.
We will reanalyze the spin ladder system with four spin cyclic exchange.
M\"uller {\it et al} and L\"auchli {\it et al} insisted that 
there is a phase transition from a gapped phase (rung-singlet phase)
to another gapped phase (staggered dimer phase).
Hikihara {\it et al} discussed that the quintet gap 
is quit small but it seems to be finite.
Honda {\it et al} insist that 
there is a phase transition from a gapped phase (rung-singlet phase) 
to a gapless phase.
But these results are not compatible.
At least one interpretation is mistaken.

We think that Honda {\it et al} found the edge state,
since they used DMRG method with the open boundary condition.
They only calculated the energy gap between the singlet state and 
the triplet state.
So, we calculate energy gaps between the singlet and triplet,
and between the singlet and the quintet.
In this paper, we try to check the consistency
between the second order phase transition and DMRG data
using a new method.
We try to calculate a correlation length quantitatively
and to discuss the finite size effect.

\section{SPIN HAMILTONIAN AND FOUR SPIN CYCLIC EXCHANGE}

\begin{figure}
\begin{center}
 \epsfig{file=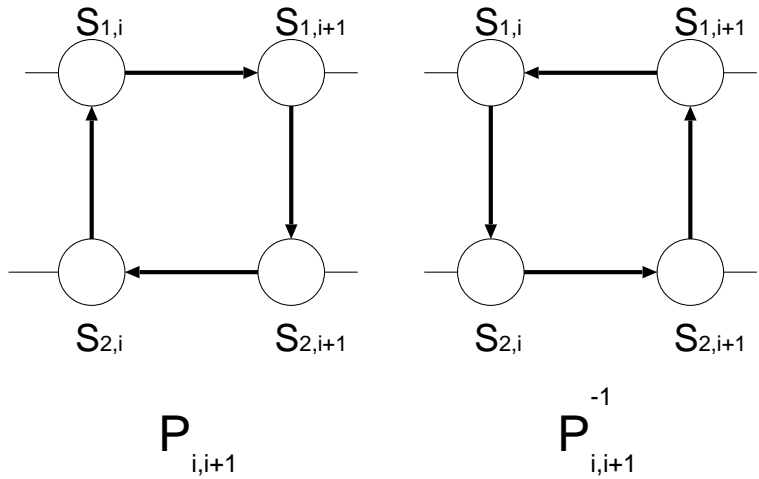,width=10cm}
 \caption{
Four spin cyclic exchange on the two-leg ladder. 
}
\label{fig:4spin-ladder2}
\end{center}
\end{figure}

In this paper, we discuss the following Hamiltonian,
\begin{eqnarray}
H &=&J_{leg} \sum_{n} \sum_{\alpha=1,2} 
{\bf S}_{\alpha,n} \cdot {\bf S}_{\alpha,n+1}
   + J_{rung} \sum_{n} {\bf S}_{1,n} \cdot {\bf S}_{2,n} \nonumber \\
   & &+ J_{ring} \sum_i \left( P_{i,i+1} + P_{i,i+1}^{-1} \right)
\label{eqn:4-s-h}
\end{eqnarray}
where $P,P^{-1}$ are four spin cyclic exchange operators
(see Fig.\ref{fig:4spin-ladder2}),
$\alpha=1,2$ are indexes of spin chains 
(see Fig.\ref{fig:4spin-ladder}),
$J_{rung}$ is an interchain coupling, 
$J_{leg}$ is an intrachain coupling,
and $J_{ring}$ is a four spin cyclic exchange coupling.

This four spin cyclic exchange operator can be expressed using spin operators
as follows,
\begin{eqnarray}
P_{i,i+1}+P^{-1}_{i,i+1}  = & & 
\frac{1}{4} + 
{\bf S}_{1,i}{\cdot}{\bf S}_{1,i+1}+{\bf S}_{2,i}{\cdot}{\bf S}_{2,i+1}+{\bf S}
_{1,i}{\cdot}{\bf S}_{2,i+1} \nonumber \\
& &
+{\bf S}_{2,i}{\cdot}{\bf S}_{1,i+1}+{\bf S}_{1,i}{\cdot}{\bf S}_{2,i}+
{\bf S}_{1,i+1}{\cdot}{\bf S}_{2,i+1} \nonumber \\
& &
+4{\{}({\bf S}_{1,i}{\cdot}{\bf S}_{2,i})({\bf S}_{1,i+1}{\cdot}{\bf S}_{2.i
+1}) \nonumber \\
& &
+({\bf S}_{1,i}{\cdot}{\bf S}_{1,i+1})({\bf S}_{2,i}{\cdot}{\bf S}_{2,i+1}) 
\nonumber \\
& &
-({\bf S}_{1,i}{\cdot}{\bf S}_{2,i+1})({\bf S}_{2,i}{\cdot}{\bf S}_{1,i+1}){
\}}.
\end{eqnarray}
This interaction have been known from old times \cite{Herring,Thouless}.

Four spin coupling terms, such as
$\left( {\bf S}_{\alpha,i} \cdot {\bf S}_{\beta,j} \right) 
\left( {\bf S}_{\gamma,k} \cdot {\bf S}_{\sigma,l} \right)$,
are derived from
fourth-order perturbation expansions
in the strong coupling limit
of the Hubbard model\cite{Takahashi}.
It is well known that
such many body exchange interactions play important role 
in Wigner crystal and $^3$He solid.
Recently, low dimensional quantum spin systems with these terms  
are well studied.
In $\frac{1}{4}J_{rung} > J_{ring}=J_{leg}$ case, 
this model can be solved exactly
using matrix product method \cite{K-M}.
On this region the ground state is a product of rung-singlet
\cite{K-M}\cite{B-M-M-N-U}.

\section{OVERVIEW OF EDGE STATES}

It is well known that
in finite size Haldane systems, under the open boundary condition, 
quasi four-hold degenerate states appear.
This was pointed out by Kennedy \cite{K},
who calculated the ground state and some lower excitation energies
numerically.

In the Haldane system, 
such as valence-bond solid (VBS) states\cite{A-K-L-T},
a ground state is unique in the infinite system. 
But under the open boundary condition,
it is four-hold degenerate, which is between the ground state and
triplet excitations, and other states, for example quintet excitations, 
have a finite energy gap.

On the other hand, 
in the finite system, under the open boundary condition,
it is quasi four-hold degenerate, which consists of the ground state
and triplet excitations.
This appears universally in Haldane gap systems.
These quasi energy gaps between four states decay exponentially,
as increasing system sizes.
This is called as edge states.
This can be confirmed experimentally\cite{H-K-A-H-R}.
Edge states should appear in not only $S=1$ chain
but also $S=1/2$ two-leg ladder system.

Usually we can confirm edge states investigating the ground state 
energies under the open boundary condition with higher magnetizations,
for example a quintet.
However, in the finite system, 
the energy gap between the singlet and triplet remains,
which results from the finite size effect.

Therefore, when the singlet-quintet gap is small,
it is difficult to investigate the edge states directly.

\section{NUMERICAL RESULTS ON PHASE TRANSITION}

In this section we present our numerical results
for low-lying levels of the model for various couplings and system sizes,
which are following as,
\begin{itemize}
 \item singlet: total $S_z=0$ ;  parity=even
 \item triplet: total $S_z=0,\pm 1$  ; parity=odd
 \item quintet: total $S_z=0,\pm 1, \pm 2$  ; parity=even.
\end{itemize}
Here we consider the parity about the inversion symmetry 
for the leg direction.
In the commensurate region, the singlet is the ground state.

Now we study $J_{leg}=J_{rung}=1$ case with verifying $J_{ring}$.
We consider spin gaps between the singlet and the triplet
and between the singlet and the quintet.
\begin{eqnarray}
\Delta E_{st}  = E \left( triplet \right) - E \left( singlet \right), \\
\Delta E_{sq} = E \left( quintet \right) - E \left( singlet \right). 
\end{eqnarray}

We calculate spin gaps, using a finite size DMRG algorithm 
with open boundary condition.
We use the maximum system size $L=112$ and $m=300$
in this paper,
where $m$ is the number of the bases for the truncated left-half system.

\subsection{RATIO OF ENERGY GAPS}

Here we consider a ratio of the energy gap of the triplet-singlet
and the energy gap of the quintet-singlet,
\begin{equation}
f \left( L, J_{ring} \right) \equiv \frac{\Delta E_{sq}}{\Delta E_{st}}.
\end{equation}

\begin{figure}
\begin{center}
 \epsfig{file=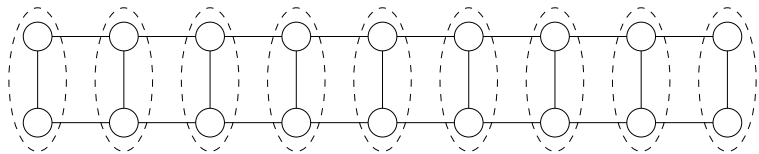,width=10cm}
 \caption{
A rung singlet state 
which consists of two spins ${\bf S}_{1,i}, {\bf S}_{2,i}$
enclosed by a dotted line.
 }
\label{fig:rung-singlet}
\end{center}
\end{figure}

\begin{figure}
\begin{center}
 \epsfig{file=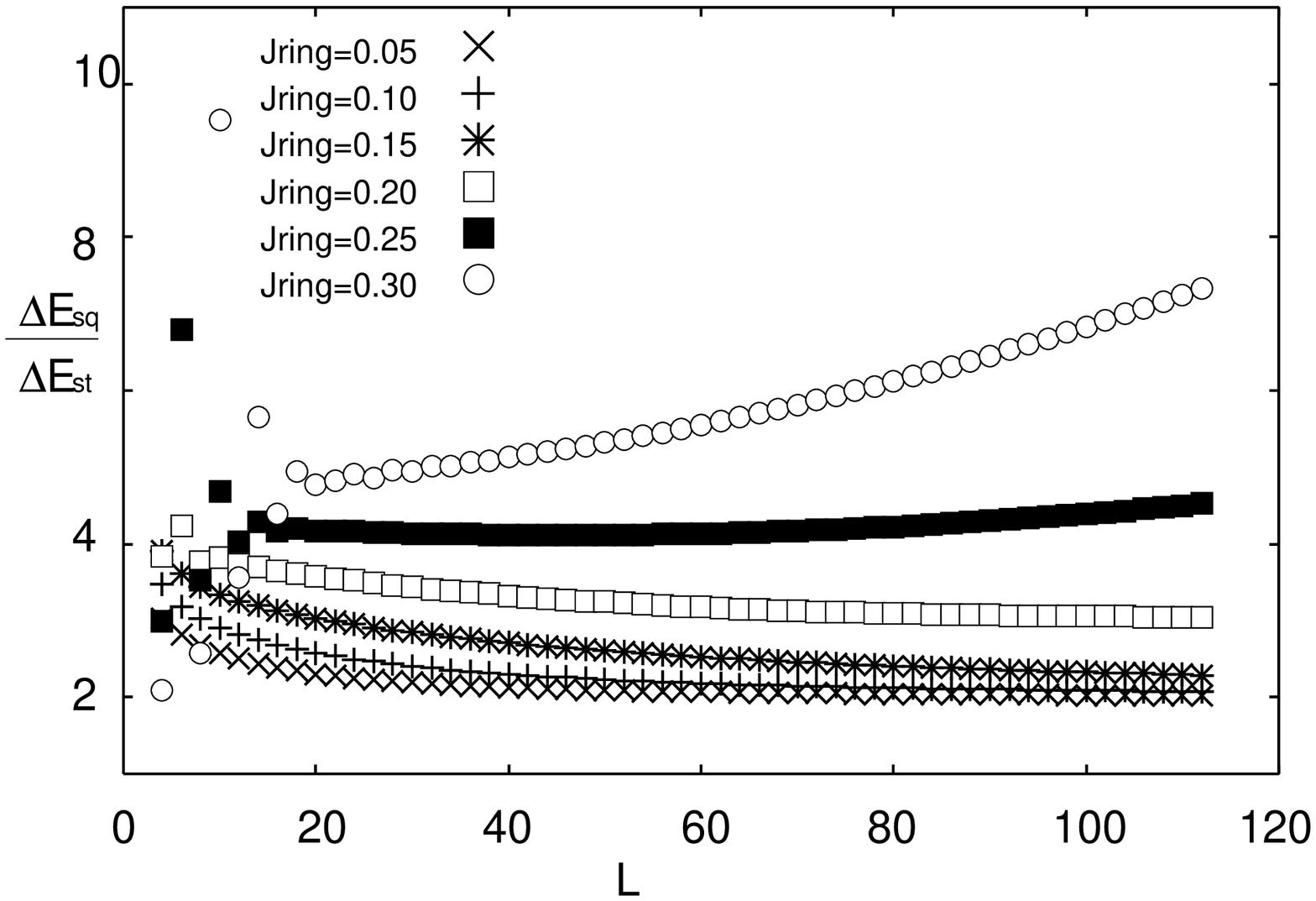,width=10cm}
 \caption{
Ratios of energy gaps
as a function of the system size $L$ 
}
\label{fig:ratio-L}
\end{center}
\end{figure}

In a normal rung-singlet phase (see Fig.\ref{fig:rung-singlet}),
this ratio is about 2, based on standard magnon picture. 
If the interaction of magnons is repulsive,
then this value should be 2 in the infinite system,
and it approach 2 from a large value
increasing the system size in the finite system.   
In this phase, the triplet excitation state has one magnon, 
and the quintet excitation state has two magnons. 
If two magnons are independent, this ratio is equal 2. 
This is true in enough large systems (see Fig.\ref{fig:ratio-L}).

In a massless phase in a finite system under open boundary condition, 
excited energies are given as
\begin{equation}
\Delta E (L) = \frac{ \pi v }{L}x_s
\end{equation}
where $x_s$ is a surface critical exponent,
$v$ is a Fermi velocity, and $L$ is a system size based on CFT.
If a phase is massless,
then this ratio converge to a finite value $x'_s/x_s$.
In reality, there remains the finite size effect from
logarithmic correction(see Appendix) 

In a staggered dimer phase, $\Delta_{st} \rightarrow 0$ and 
$\Delta_{sq} \rightarrow finite$, so $f \rightarrow \infty$.

So we can distinguish a massless phase from some degenerate phases
in large $J_{ring}$ region, as follows,
\begin{itemize}
 \item Massless phase   \ \ \ \ \ \ \ \ \ \ : $f$ $\rightarrow$ converge (finite)
 \item Staggered dimer phase : $f$ $\rightarrow$ diverge
\end{itemize}

\begin{figure}
\begin{center}
 \epsfig{file=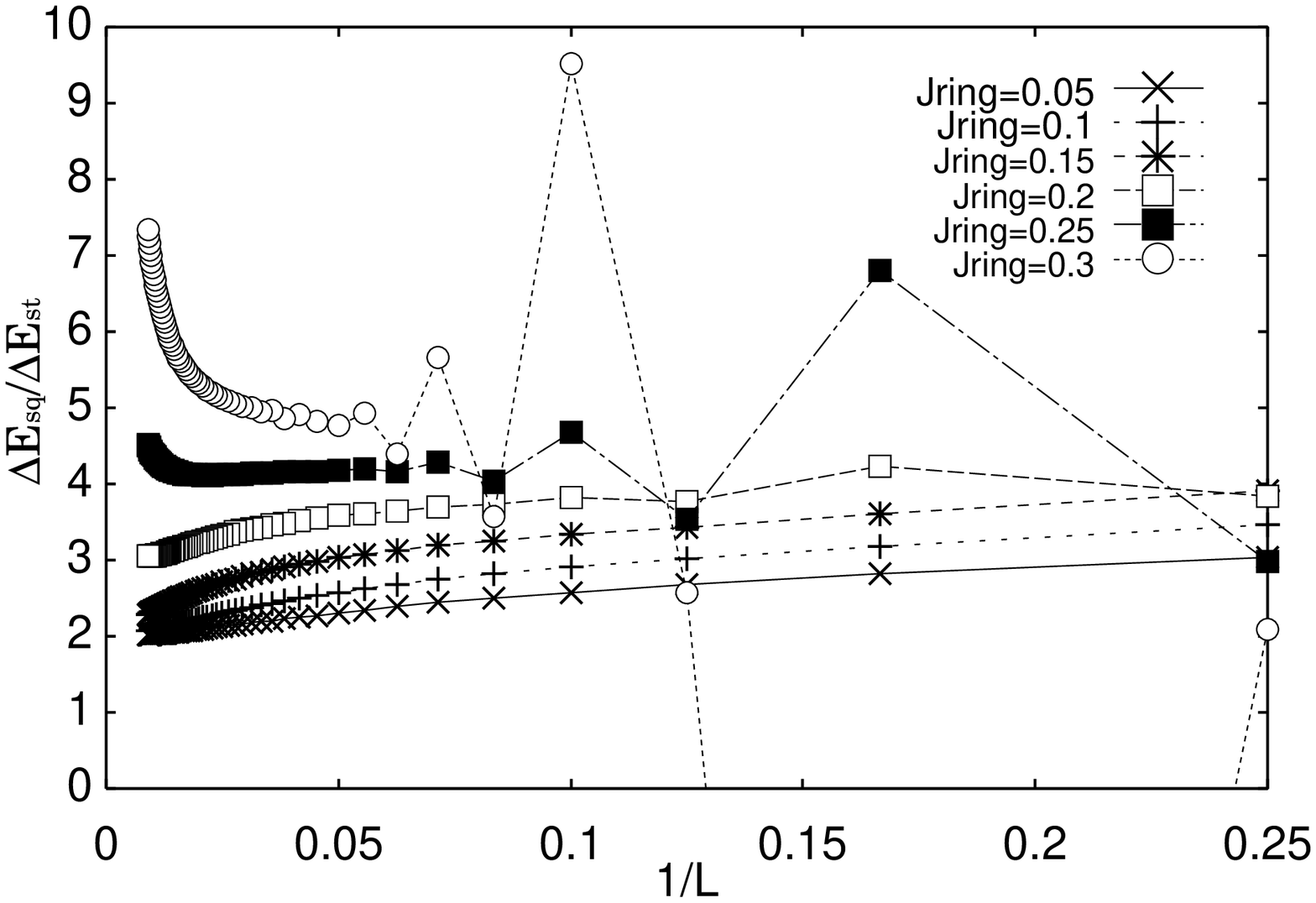,width=10cm}
 \caption{
Ratios of energy gaps
as a function of the system size $1/L$ 
 }
\label{fig:ratio-invL}
\end{center}
\end{figure}

We find that this ratio tend to diverge in large $J_{ring}$ region
(see Fig.\ref{fig:ratio-L},\ref{fig:ratio-invL}).
This shows that $f \rightarrow 2$ in the small $J_{ring}$ region,
and $f \rightarrow \infty $ in the large $J_{ring}$ region.
So we can deny that there exists an extended massless region
in the large $J_{ring}$ region.
And it shows that the interaction of magnons is repulsive
in the rung-singlet phase. 
We can understand $f \rightarrow 2$ to consider that
the rung-singlet phase appears.
And we can interpret $f \rightarrow \infty$ as
$\Delta_{st} \rightarrow 0$ and $\Delta_{sq} \rightarrow finite$.
Considering $SU(2)$ symmetry, we can conclude that the edge state 
appears.

Here we consider the function $f$.
This phase transition is second order,
where the critical point is massless,
so we can expect that this function behave
\begin{itemize}
 \item $J_{ring} < J_{ring}^c$ : $f \rightarrow $finite $\sim$ 2
 \item $J_{ring} = J_{ring}^c$ : $f \rightarrow x_s(sq)/x_s(st)$ $\sim$ finite
 \item $J_{ring} > J_{ring}^c$ : $f \rightarrow \infty$
\end{itemize}
as increasing system size $L$,
where $x_s (st),x_s(sq)$ are scaling dimensions.
This function is independent of system size $L$
at $J_{ring}=J_{ring}^c$.
Therefore, when we plot this function for $J_{ring}$,
we can expect that this function behaves as
the traditional finite size scaling function $L\Delta E$.
This is based on a phenomenological
renomalization group analysis \cite{F-B}.

\begin{figure}
\begin{center}
 \epsfig{file=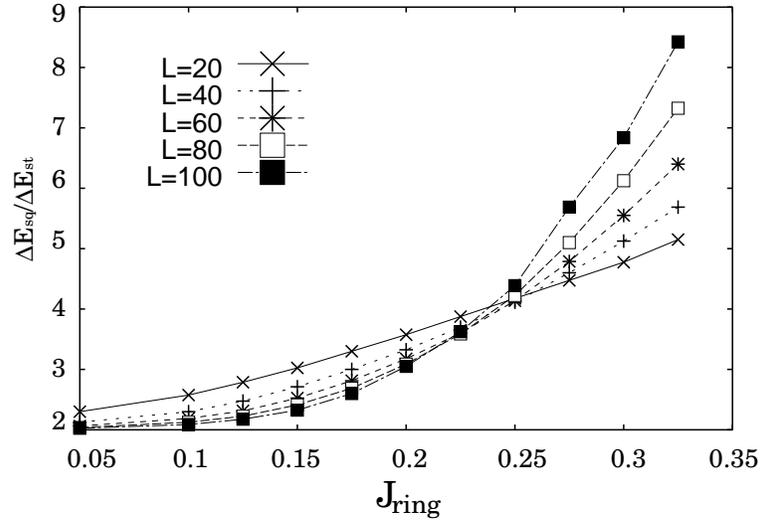,width=10cm}
 \caption{
Ratios of energy gaps
as a function of the system size $J_{ring}$ 
 }
\label{fig:ratio-jring}
\end{center}
\end{figure}

In Fig.\ref{fig:ratio-jring}
we define $J_{ring}^{cross} \left( L \right)$ as

\begin{equation}
f \left( L+20, J_{ring}^{cross} \left( L+10 \right) \right)=
f \left( L, J_{ring}^{cross} \left( L+10 \right) \right)
\end{equation}

\begin{figure}
\begin{center}
 \epsfig{file=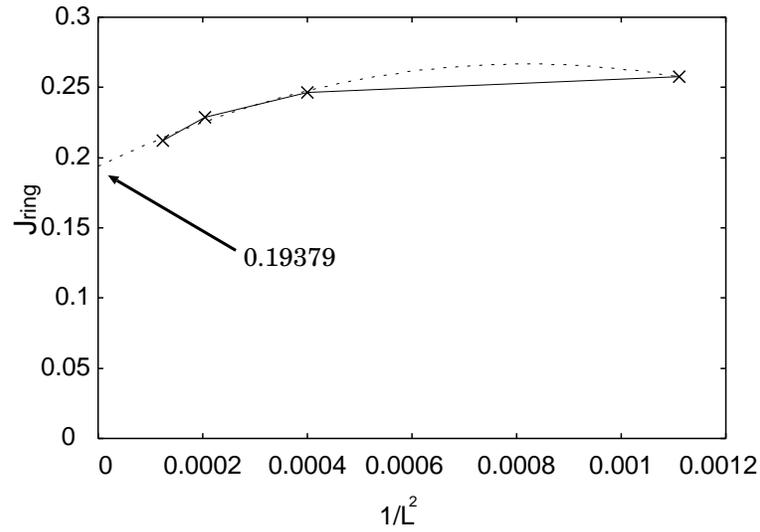,width=10cm}
 \caption{
$J_{ring}^{cross}$ as function of the system size $1/L$.
 }
\label{fig:ext-cross}
\end{center}
\end{figure}

Then, we extrapolate $J_{ring}^{cross}$ as follows,
\begin{equation}
J_{ring}^{cross} \left( L \right) 
= J_{ring}^{cross} \left( \infty \right) 
+ a \frac{1}{L^2} + b \frac{1}{L^4}
+ \mbox{higher order}
\end{equation}
Here we neglect higher order terms,
so we obtain $J_{ring}^{cross} \left( \infty \right)=0.19379$
(see Fig.\ref{fig:ext-cross}).

\subsection{THE CORRELATION LENGTH}

In this section, we discuss the correlation length
in the rung-singlet phase.
We determine the correlation length $\xi$ 
according to the following formula,
\begin{equation}
f \left( L , J_{ring} \right) = 
f \left( L=\infty, J_{ring} \right)  +a K_0 \left( L / \xi \right)
\end{equation}
where $a$ is an amplitude, 
$K_0$ is the modified Bessel function,
and $f \left( L=\infty, J_{ring} \right) $ is about 2.
As is well known, the $d$ dimensional quantum system can be
mapped onto $d+1$ dimensional classical system\cite{S-T}.
If the one-dimensional quantum system has the energy gap,
it is called as massive, 
and its correlation function becomes of 
the $1+1$ Ornstein-Zernicke form,
that is the modified Bessel function,
\begin{equation}
\int \frac{ \exp \left[ i {\bf q} \cdot {\bf l} \right] }
{ q^2 + \xi ^{-2} } d^dq \propto
\left\{ \begin{array}{ccc}
     \exp \left[ - \left| {\bf l } \right| / \xi\right] & : d=1 &\\
      K_0 \left(  \left| {\bf l} \right| / \xi\right) & : d=2 &.
        \end{array} \right. 
\end{equation}
This form is supported by Quantum Monte Carlo simulation \cite{Nomura}
and DMRG calculation \cite{W-H}

Using the nonlinear least squares method,
we determine $f(L=\infty,J_{ring}),a,\xi$. 
We obtain $1/\xi$ as a function of $J_{ring}$,
fitting for $L=20 \sim 112$.
This figure shows clearly that 
the $1/\xi$ is proportionate to $J_{ring}$.
Note that in the case of $J_{ring}=0.175$,
$\xi$ is larger than the system size $L$.
The correlation length diverges
at the second order phase transition point.
Our result is consistent with this fact.

\begin{figure}
\begin{center}
 \epsfig{file=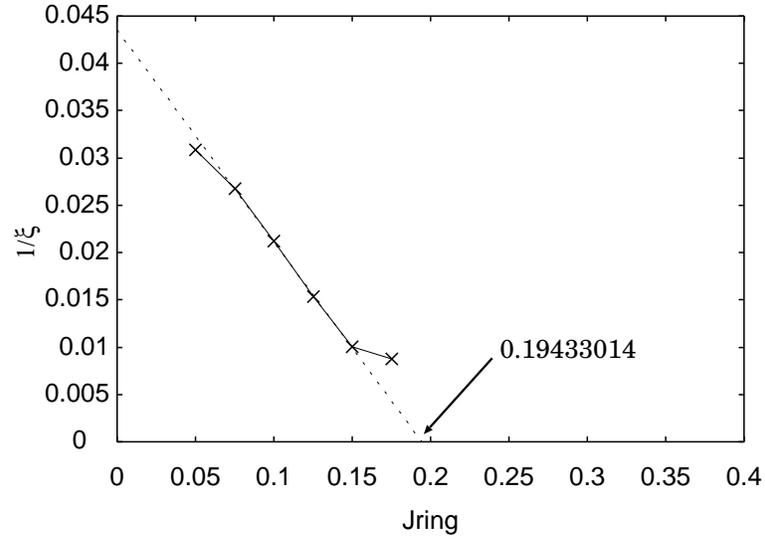,width=10cm}
 \caption{
The  correlaiton length as a function of $J_{ring}$ is 
computed using the modified Bessel function.
The dot line was computed using a least-squares method.
 }
\label{fig:correlation1}
\end{center}
\end{figure}

Near the critical point described as $k=2$ $SU(2)$ WZW model,
$1/\xi$ behave
\begin{equation}
\frac{1}{\xi} \propto \left| J_{ring} - J_{ring}^{c} \right| ,
\end{equation}
where $J_{ring}^c$ is $J_{ring}$ at the critical point.
Using the method of least squares,
we obtain the $\xi^{-1}=0$ at $J_{ring}^c=0.1943$
(see Fig. \ref{fig:correlation1}).
This value is close to previous results \cite{H-N,L-S-T}.

\subsection{SCALING DIMENSION}

In previous section, we discussed the divergence of the correlation
length. 
In this section, returning to the CFT analysis
with the periodic boundary condition,
we discuss a scaling dimension independently.
In our previous paper \cite{H-N},
we obtained a scaling dimension $x=\frac{3}{8}$
thus we have calculated $k=2$ WZW type central charge $c=\frac{3}{2}$.
But this model, 
level $k=2$ $SU(2)$ WZW model,
has another scaling dimension $x=1$.
This scaling dimension is related with the energy gaps
(the correlation length $\xi^{-1} \propto \Delta E$)
near the critical point,
\begin{equation}
\Delta E \propto \left| J_{ring} - J_{ring}^c \right| ^{\eta} 
\end{equation}
where $\eta =1$.
M\"uller {\it et al} \cite{M-V-M} got this critical exponent,
using the cluster expansion and Pad\'e approximation.
This is based on rung-dimer limit $J_{rung} \rightarrow \infty$.
However we are interested in $J_{leg}=J_{rung}$ case.
Now we consider a scaling dimension based on CFT.
The relation between an excitation energy and a scaling dimension
,under the periodic boundary condition,
is 
\begin{equation}
E_i - E_0 = \frac{2 \pi v}{L}x_i
\end{equation}
where $L$ is a systems size, 
$v$ is a Fermi velocity in the system and 
$x_i$ is a scaling dimension.
In addition to the ground state energy and the excitation energy,
we need to obtain $v$ numerically.
This velocity is obtained from
\begin{equation}
v \left( L \right) = \frac{L}{2 \pi}
\left[ 
E \left( q=\frac{2 \pi}{L} \right) - E_{gs}  
\right]
\end{equation}
where $q$ is a wave number.
Then we extrapolate $v(L)$ as follows,
\begin{equation}
v \left( L \right) = v_{\infty} + a \frac{1}{L^2} 
+ b \left( \frac{1}{L^2} \right)^2
+ \mbox{higher order}
\end{equation}
We neglect higher order terms and use $v_{\infty}$.

\begin{figure}
\begin{center}
 \epsfig{file=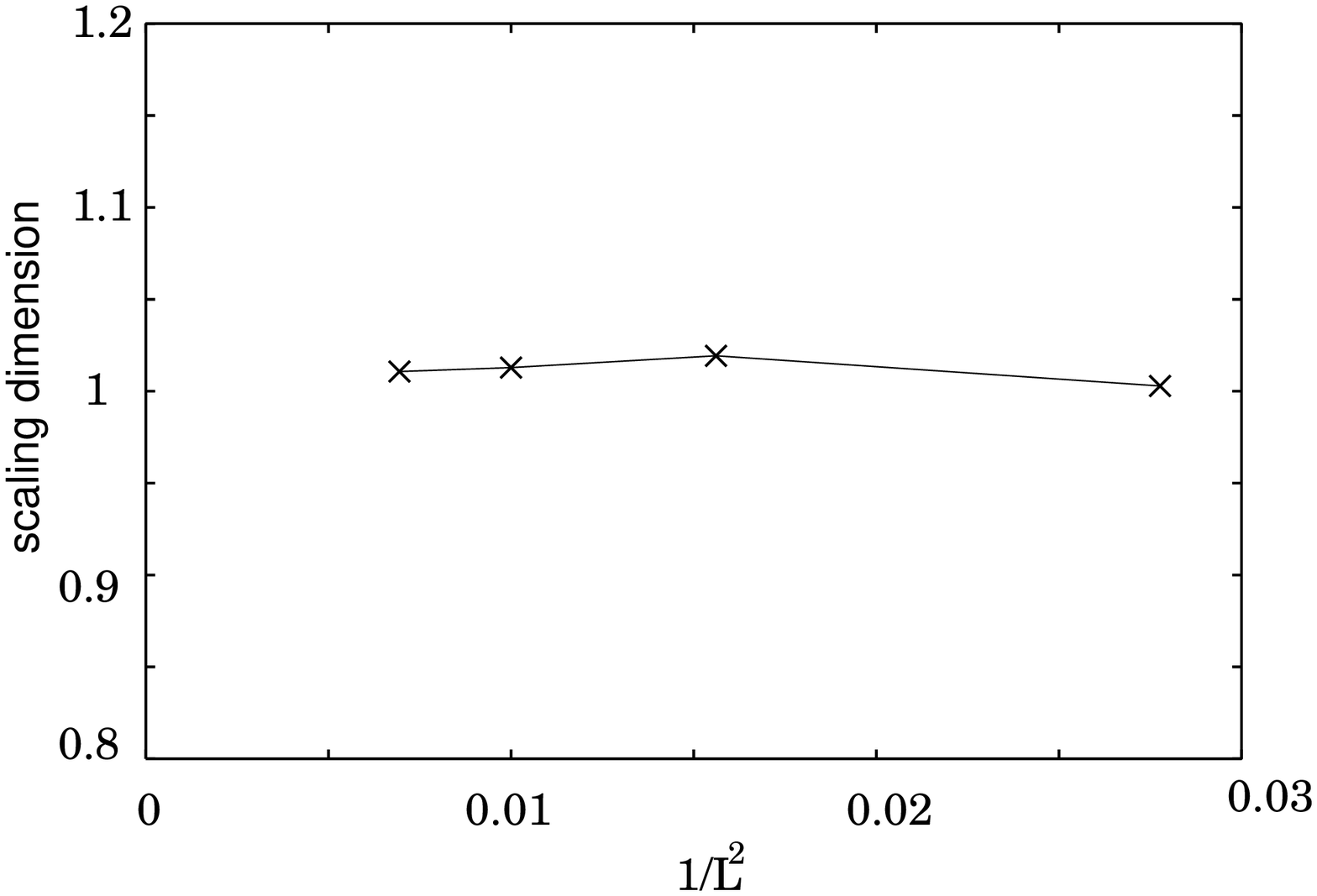,width=10cm}
 \caption{
Size dependence of the scaling dimension $x=1$
 after removing logarithmic correction.
 }
\label{fig:sdim}
\end{center}
\end{figure}

Unfortunately there are logarithmic corrections 
from the marginal operator
in $k=2$ $SU(2)$ WZW model.
So we must remove corrections,
when we calculate scaling dimensions.
After removing them with the method in Appendix,
a scaling dimension is almost independent on the system size
at the critical point $J_{ring}=0.192$
(see Fig.\ref{fig:sdim}).
This result is consistent with previous results \cite{H-N,M-V-M}.

\section{CONCLUSIONS}

In this paper
we have investigated the ordered state of a $S=\frac{1}{2}$ quantum 
spin two-leg ladder system with the cyclic four spin exchange
using DMRG method 
in the open boundary condition
and an exact diagonalization method
in the periodic boundary condition.

Generally speaking, it is very difficult to determine
whether some value is zero or not numerically,
especially near the critical point,
where the correlation length becomes large 
compared with the system size.

Considering the ratio of energy gaps of the singlet-triplet and 
the singlet-quintet,
we find that the singlet-triplet is degenerate and
the singlet-quintet is not degenerate after the phase transition.
This result supports that the staggered dimer phase appears.
In the staggered dimer phase with the open boundary condition,
since the edge state appears, 
the singlet-triplet energy gap decays exponentially.
On the other hand, the singlet-quintet gap is finite in this phase. 

We have also investigated the correlation length
near the critical point.
That result is consistent with previous results
\cite{H-N,M-V-M,L-S-T}.

Honda {\it et al} \cite{H-H} concluded that 
the massless phase appeared,
using DMRG with the open boundary condition.
But we can consider that they found the small edge state. 
We obtain the unified interpretation which 
is consistent with previous results
\cite{H-N,M-V-M,H-H,H-M-H,L-S-T}.

This problem is related to electron systems on a two-leg ladder.
Our result is consistent with Tsuchiizu {\it et al} \cite{T-F}. 
They discussed the ground state phase diagram of half filled 
two-leg Hubbard ladder with inter-site Coulomb repulsions and exchange
coupling.
They insisted that the universality class of  
the phase transition between the rung-singlet phase
and the staggered dimer phase is $c=3/2$ CFT or the first order 
phase transition.  
Our result is consistent with their result.

According to Hikihara {\it et al} this model has 
a self-duality ($J_{ring}=J_{leg}/2$),
which is between a spin and a vector chirality \cite{H-M-H}.
In the case that the system has duality and there exists
a phase transition, 
if a phase transition point is not the self-dual point,
there must be another phase transition.
Now Hamiltonian is not invariant 
for the duality transformation exactly.
But we can expect that there is another phase transition
reflected in the duality transformation.
This should be a future problem.

\section{ACKNOWLEDGMENTS}

Authors are grateful to A. Kitazawa for fruitful discussions.
This work is partly supported by Grant-in-Aid for Scientific
Research NO. 12000039 from the Ministry of Education, Science and
Culture, Japan.
A part of numerical calculations in the paper was
based on the package TITPACK ver 2.0 by H. Nishimori.

\appendix

\section{WESS-ZUMINO-WITTEN MODEL AND LOGARITHMIC CORRECTION 
\cite{gins,A-H,A-G-S-Z}}

The minimal conformal field theory with the smallest spectrum
containing currents obeying the Kac-Moody algebra with 
central charge $c$ is the Wess-Zumino-Witten non-linear $\sigma$
model, with topological coupling constant $k$.
Thus the relation between $c$ and $k$ is 
\begin{equation}
 c=\frac{k \dim G}{k+\frac{C_A}{2}},
\end{equation}
where $\dim G$ is dimension for an arbitrary representation 
and $\frac{C_A}{2}$ is called as the dual Coxeter number.
In the present case, since the system has SU(2) symmetry,
we have $\dim G = 3$, and $\frac{C_A}{2} = 2$.

Now we think of the case with the SU(2) symmetry.
In this case, primary fields can be classified according to 
their left and right moving spin. There are operators with 
$s_L=s_R=0, \frac{1}{2},...,\frac{k}{2}$.
Their scaling dimensions are 
\begin{equation}
 x=\frac{2s_L \left( s_L + 1 \right)}{2+k}.
\end{equation}
The operators with $s_L$ half-odd integer (integer) are odd (even)
under translation by one site; thus they correspond to states
with momentum $\pi$ (zero).

Here we think of the case of the $k=2$ SU(2) WZW model.
In the case of $s_L=s_R=\frac{1}{2}$,
which forms $\frac{1}{2} \otimes \frac{1}{2} = 0 \oplus 1$,
the scaling dimension is $x=\frac{3}{8}$.
In the case of $s_L=s_R=1$ which forms
$1 \otimes 1 = 0 \oplus 1 \oplus 2$,
the scaling dimension is $x=1$.
This is related to the critical exponent of a  energy gap.

Besides Kac-Moody primary operators,
there is a marginal operator for all $k$,
namely $ \vec{J}_L \cdot \vec{J}_R$
(which is a primary field with respect to the Virasoro algebra).
Current operator $\vec{J}_L$ 
has the conformal weight $(h,\bar{h})=(1,0)$ and
$\vec{J}_R$ has the conformal weight $(h,\bar{h})=(0,1)$.
Thus the $ \vec{J}_L \cdot \vec{J}_R$ operator has the conformal weight 
$(h,\bar{h})=(1,1)$
which has the scaling dimension $x=h+\bar{h}=2$
and the conformal spin 0, i.e. wave number $q=0$.

Current operators $ \vec{J}_L$, $\vec{J}_R$ themselves
have conformal spin $ \pm 1$, corresponding to the wave number
$q= \pm 2 \pi/L$, thus they are related with the spin wave velocity.


According to the non-Abelian bosonization \cite{Witten}, 
in SU(2) symmetric gapless system,
the excitation energies (for $q= 0$ states)
in the periodic boundary condition, 
including log corrections, are 
\begin{equation}
 \Delta \mbox{E} = \mbox{E}_i - \mbox{E}_0 \approx \frac{2 \pi v}{L}
  \left( x_i - \frac{ \left\langle {\bf S}_L \cdot {\bf S}_R \right\rangle}{ \ln L}
  \right).
\end{equation}
where $L$ is a system size, and $v$ is a spin wave velocity.
Here ${\bf S}={\bf S}_L+{\bf S}_R$ is total spin which is a
conserved quantity.
Note that
\begin{eqnarray}
 {\bf S}_L \cdot {\bf S}_R&=& \frac{1}{2} \left( {\bf S}_L + {\bf S}_R \right)^2
                             -\frac{1}{2} {\bf S}_L^2 - \frac{1}{2} {\bf S}_R^2 \\
                           &=&\frac{1}{2} s(s+1) - \frac{1}{2} s_L \left(s_L +1 \right) 
                                 - \frac{1}{2} s_R \left(s_R +1 \right).
\end{eqnarray}
In the case of $s_L=s_R=1$,
since $s_L \otimes s_R$=$1 \otimes 1$=$0 \oplus 1 \oplus 2$,
it forms a quintet $s=2$ triplet $s=1$ and a singlet $s=0$.
The states $s_L=s_R=1$ correspond to a wave vector $q=0$.

First $s=0$(singlet) case,
\begin{equation}
 \left\langle singlet \mid {\bf S}_L \cdot {\bf S}_R \mid singlet
\right\rangle = -2
\end{equation}
Second $s=1$(triplet) case,
\begin{equation}
 \left\langle triplet \mid {\bf S}_L \cdot {\bf S}_R \mid triplet
\right\rangle = -1
\end{equation}
Third $s=2$(quintet) case,
\begin{equation}
 \left\langle quintet \mid {\bf S}_L \cdot {\bf S}_R \mid quintet
\right\rangle = 1
\end{equation}
Therefore we obtain 
\begin{equation}
 \Delta \mbox{E}(s=2) =
 \mbox{E}_i - \mbox{E}_0 \approx \frac{2 \pi v}{L}
 \left( x_i - \frac{1}{ \ln L}
 \right),
\end{equation}
and
\begin{equation}
 \Delta \mbox{E} (s=1) =
 \mbox{E}_i - \mbox{E}_0 \approx \frac{2 \pi v}{L}
 \left( x_i + \frac{1}{ \ln L}
 \right),
\end{equation}
and
\begin{equation}
 \Delta \mbox{E} (s=0) =
 \mbox{E}_i - \mbox{E}_0 \approx \frac{2 \pi v}{L}
 \left( x_i + 2 \frac{1}{ \ln L}
 \right),
\end{equation}
thus we can remove logarithmic corrections of energy gap.
\begin{displaymath}
 \Delta \mbox{E}(s=2) + \Delta \mbox{E}(s=1) = 2 \frac{2 \pi v}{L} x_i,
\end{displaymath}
or
\begin{equation}
 x_i = \frac{ L }{4 \pi v} \left[ \Delta \mbox{E}(s=2)
                                + \Delta \mbox{E}(s=1) \right].
\end{equation}

\newpage

\begin{thebibliography}{99.}

\addcontentsline{toc}{section}{References}


\bibitem{D-R} E. Dagotto and T. M. Rice, Science {\bf 271} 618 (1996),
E. Dagotto, Rep. Prog. Phys. {\bf 62}, 1525 (1999)

\bibitem{Haldane} F. D. M. Haldane, Phys. Rev. Lett {\bf 50} 1153 (1983)
; Phys Lett. {\bf 93 A}, 464 (1983)

\bibitem{D-R-S} E. Dagotto, J. Riera, and D. Scalapino, Phys. Rev. B {\bf45} 5744 (1992) 

\bibitem{N-T} A. A. Nersesyan and A. M. Tsvelik, Phys. Rev. Lett {\bf 78} 3939
(1997) 


\bibitem{N-B-K-W-G}T. S. Nunner, P. Brune, T. Kopp, M. Windt, and
	M. Gr\"uninger, Phys. Rev. B {\bf 66} 180404(R) (2002)

\bibitem{S-K-U} K. P. Schmidt, C. Knetter and G
	S. Uhrig,Europhys. Lett. {\bf 56}, 877 (2001)

\bibitem{B-L-S-U} A. B\"uhler, U. L\"ow, K. P. Schmidt and G. S. Uhrig
	cond-mat/0210669 (2002)

\bibitem{S-M-U} K. P. Schmidt, H. Monien and G. S. Uhrig,
	cond-mat/0211429 (2002)
	
\bibitem{M-K-E-B-M-1}
       M. Matsuda, K. Katsumata, R. S. Eccleston, S. Brehmer and
         H-J. Mikeska, Phys. Rev. B {\bf 62} 8903 (2000)

\bibitem{M-K-E-B-M-2}
         M. Matsuda, K. Katsumata, R. S. Eccleston, S. Brehmer and
         H-J. Mikeska, J. Appl. Phys {\bf 87} 6271 (2000).

\bibitem{C-H-A-P-F-M-C-F}
         R. Coldea, S. M. Hayden, G. Aeppli, T. G. Perring, C. D. Frost,
         T. E. Mason, S. -W. Cheong, and Z. Fisk,
         Phys. Reb. Lett, {\bf 86} 5377 (2001) 

\bibitem{H-K-W} Y. Honda, Y. Kuramoto, and T. Watanabe, Phys. Rev. B,
                {\bf 47} 11329 (1993) 

\bibitem{S-H} T. Sakai and Y. Hasegawa Phys.Rev, B {\bf 60} 48 (1999)

\bibitem{N-T-H-O-S} A. Nakasu, K. Totsuka, Y. Hasegawa, K. Okamoto
                    and T. Sakai, J. Phys. Condens. Matter {\bf 13}
                    7421 (2001)

\bibitem{B-K-T} Z. L. Berezinskii, Zh. Eksp. Teor. Fiz, {\bf 59}, 907 
(1970) (Sov. Phys. JETP, {\bf 32}, 493 (1971))
; Zh. Eksp. Teor. Fiz, {\bf 61}, 1144, (1971)
(Sov. Phys. JETP, {\bf 34}, 610 (1972)),
J. M. Kosterliz and D. J. Thouless, J. Phys. C, {\bf 6}, 1181 (1973),
J. M. Kosterliz, J. Phys. C, {\bf 7}, 1046 (1974)

\bibitem{H-N} K. Hijii and K. Nomura, Phys. Rev. B, {\bf 65}, 104413 
(2002)


\bibitem{gins} See,for a review,P. Ginsparg :
Fields, Strings and Critical Phenomena 
               ed E. Brezin and J. Zinn-Justin (1989),
             and references therein.  

\bibitem{M-V-M} M. M\"uller, T. Vekua, and H.-J. Mikeska,
                Phys. Rev. B, {\bf 66}, 134423 (2002)

\bibitem{L-S-T}
A. L\"auchli, G. Schmid, and M. Troyer, cond-mat/0206153 (2002)

\bibitem{H-H} Y. Honda, and T. Horiguchi, cond-mat/0106426

\bibitem{DMRG} S. R. White, Phys. Rev. Lett, {\bf 69}, 2863 (1992)
; Phys. Rev. B, {\bf48}, 10345 (1993)


\bibitem{H-M-H}
T. Hikihara, T. Momoi, and X. Hu, cond-mat/0206102 (2002)

\bibitem{K-M} A. K. Kolezhuk and H.-J. Mikeska, Int. J. Mod. Phys. B
              {\bf 12}, 2325 (1998)


\bibitem{Herring} C. Herring, Rev. Mod. Phys. {\bf 34}, 631 (1962)
           ; {\it Magnetism}, edited by G. T. Rado and H. Suhl 
           (Academic, New York, 1966), Vol. 2. B

\bibitem{Thouless} D. J. Thouless, Proc. Phys. Soc. {\bf 86}, 893 (1965) 

\bibitem{Takahashi} M. Takahashi, J. Phys. C, {\bf 10} 1289 (1977).
                     ; {\it Thermodymamics of One-Dimensional Solvable Models}, Cambridge Univ Press

\bibitem{B-M-M-N-U} S. Brehmer, H.-J. Mikeska, M. M\"uller, N. Nagaosa, and S.
Uchida, Phys. Rev. B {\bf 60}, 329 (1999)


\bibitem{K} T. Kennedy, J. Phys. Condens. Phys. Matter {\bf 2}
            5737 (1990)

\bibitem{A-K-L-T} I. Affleck, T. Kennedy, E. H. Lieb, and H. Tasaki, 
Phys. Rev. Lett. {\bf 59}, 799 (1987); Commun. Math. Phys. {\bf 115}, 477
(1989)

\bibitem{H-K-A-H-R} M. Hagiwara,K. Katsumata, I. Affleck, B. I. Halperin and J. P. Renard, Phys. Rev. Lett. {\bf 65} 3181 (1990)


\bibitem{F-B} M. E. Fisher and M. N. Barber, Phys. Rev. Lett. {\bf 28} 1516 (1972)


\bibitem{S-T} H. F. Trotter, Proc. Am. Math. Soc. {\bf 10} 545 (1959)
M. Suzuki, Prog. Theor. Phys. {\bf 56} 1454 (1976);
J. Stat. Phys. {\bf 43} 886 (1986)

\bibitem{Nomura} K. Nomura, Phys. Rev. B. {\bf 40} 2421 (1989)

\bibitem{W-H} S. R. White and D. A. Huse Phys. Rev. B. {\bf 48} 3844


\bibitem{T-F} M. Tsuchiizu and A. Furusaki, cond-mat/0206539
(2002)


\bibitem{A-H} I. Affleck and F. D. M. Haldane:Phys. Rev. B {\bf 36} 5291 (1987)

\bibitem{A-G-S-Z} I. Affleck, D. Gepner, H. J. Schulz, and T. Ziman,
         J. Phys. A {\bf 22} 511 (1989)

\bibitem{Witten} E. Witten, Commun.Math.Phys 92 455 (1984)


\end{thebibliography}
\end{document}